# ANALYSIS OF CITIZENS' ACCEPTANCE FOR E-GOVERNMENT SERVICES: APPLYING THE UTAUT MODEL *

Mohammed Alshehri[1], Steve Drew[1] and Rayed AlGhamdi[2]

[1]*School of ICT, Griffith University, Brisbane, Australia*
[2]*Faculty of Computing & IT - King Abdulaziz University, Kingdom of Saudi Arabia*

ABSTRACT

E-government services aims to provide citizens with more accessible, accurate, real-time and high quality services and information. Although the public sectors in Kingdom of Saudi Arabia (KSA) have promoted their e-Government services for many years, its uses and achievements are few. Therefore, this paper explores the key factors of Saudi citizens' acceptance through a research survey and by gathering empirical evidence based on the Unified Theory of Acceptance and the Use of Technology (UTAUT). Survey Data collected from 400 respondents was examined using structural equation modelling (SEM) technique and utilized AMOS tools. The study results explored the factors that affect the acceptance of e-government services in KSA based on UTAUT model. Moreover, as a result of this study an amended UTAUT model was proposed. Such a model contributes to the discussion and development of adoption models for technology.

## 1. INTRODUCTION

E-government has become a popular focus of government efforts in many countries around the world. Many of these countries have proceeded to implement and adopt their E-government systems and services. It reflects the intention for public organizations and governments to take advantage of the communications improvements made possible by the Information and Communication Technology (ICT) revolution. E-government is adopted with the purpose of improving the services and delivery provided by the government to its citizens (Imran and Gregor, 2007). E-government services systems aim to provide many benefits such as improving the processes and operations of government services and enhancing information sharing between the government and public. It also provides citizens the services in professional manner, securely, safely, conveniently, and with considerable time savings. However, the implementation of e-government services is not a simple on-line information provision, it requires a deep understanding of citizens needs and requirements and comprehensive architecture to avoid unexpected results (Kaliannan et al., 2007). E-government and Internet has made an essential change in the whole Saudi society structure, values, culture and the ways of conducting business by utilizing the potential of ICT as a tool in daily work. Therefore the ongoing development of e-government has been identified as one of the top priorities for Saudi government and all its agencies. In this paper we identify the factors that influence Saudi citizens to accept and use e-government by applying an amended UTAUT Model. UTAUT is an empirically validated model combining eight major models of technology acceptance and their extensions.

---

## 2. LITERATURE REVIEW

## 2.1 E-government Background

E-government is a new wave in the information revolution. Many governments around the world follow this phenomenon hoping to reduce costs, improve services delivery for citizens and to increase effectiveness and efficiency in the public sector. E-government represents an essential change in the whole public sector structure, values, culture and the ways of conducting business. In fact, there are many definitions for the term e-Government and differences reflect the priorities in the government strategies. Moon and Norris (2005) provides a simple definition that e-government is perceived as "means of delivering government information and service" (p.43). Isaac (2007) defined electronic government as government's use of technology, particularly Web-based Internet applications, to enhance the access to and delivery of government information and service to citizens, business partners, employees, other agencies, and government entities. Similarly, Fang (2002) defined e-government as a way for governments to use the most innovative information and communication technologies, particularly Web-based Internet applications, to provide citizens and businesses with more convenient access to government information and services, to improve the quality of the services and to provide greater opportunities to participate in democratic institutions and processes. Moreover, Carter and Belanger (2005) defined e-government services as the use of ICT to enable and improve the efficiency of the government services that are provided to citizens, employees, businesses, and agencies. According to Carter and Belanger (2005), e-government services increase the convenience and accessibility of government services and information to citizens. Nowadays, government agencies around the world are increasingly making their services available online. E-government services become especially important given its potential to reduce costs and improve service compared with traditional modes of government service delivery (Carter & Belanger, 2005).

## 2.2 The Unified Theory of Acceptance and Use of Technology (UTAUT)

Information technology acceptance and adoption research has developed several competing and complementary models each with a different set of acceptance determinants. These models have evolved over the years and came as a result of persistent efforts towards models' validation and extension that took place during the period each was presented to the research community. Most notable amongst these models are the Theory of Reasoned Action (TRA) (Ajzen & Fishbein, 1980), Theory of Planned Behaviour, (TPB) (Ajzen, 1985), Technology Acceptance Model (TAM) (Davis, 1989), Extension of the Technology Acceptance Model (TAM2) (Venkatesh and Davis, 2000), Diffusion of Innovation Model (DOI) (Rogers, 2003), and Unified Theory of Acceptance and Use of Technology (UTAUT) (Venkatesh et al., 2003). UTAUT is one of the latest developments in the field of general technology acceptance models. Like earlier acceptance and adoption models, it aims to explain user intentions to use an Information System (IS) and further the usage behavior. Venkatesh et al. (2003) created this synthesized model to present a more complete picture of the acceptance process than any previous individual models had been able to do. Eight models previously used in the IS literature were merged in an integrated model, all of which had their origins in psychology, sociology and communications. These models are the TRA, TPB, TAM, TAM2, the Motivational Model of Computer Usage (MM) (Igbaria et al., 1996), the Model of PC Utilization (MPCU) (Thompson et al., 1991), DOI and Social Cognitive Theory (SCT) (Bandura, 1977; Compeau, et al., 1999). Each model attempts to predict and explain user behaviour using a variety of independent variables. A unified model was created based on the conceptual and empirical similarities across these eight models. The UTAUT holds that four key constructs (performance expectancy, effort expectancy, social influence, and facilitating conditions) are direct determinants or predictors of usage intention and behavior (Venkatesh et. al., 2003). Gender, age, experience, and voluntariness of use are posited to mediate the impact of the four key constructs on usage intention and behavior as indicated in Figure 1. The predictors are defined as follows (Venkatesh et al. 2003, 447-453):

    1. Performance expectancy (PE): "*is the degree to which an individual believes that using the system will help him or her to attain gains in job performance.*"

    2. Effort expectancy (EE): "*is the degree of ease associated with use of the system.*"

3. Social influence (SI): "*is the degree to which an individual perceives that important others believe he or she should use the new system.*"

4. Facilitating conditions (FC): "*is the degree to which an individual believes that an organizational and technical infrastructure exists to support use of the system.*"

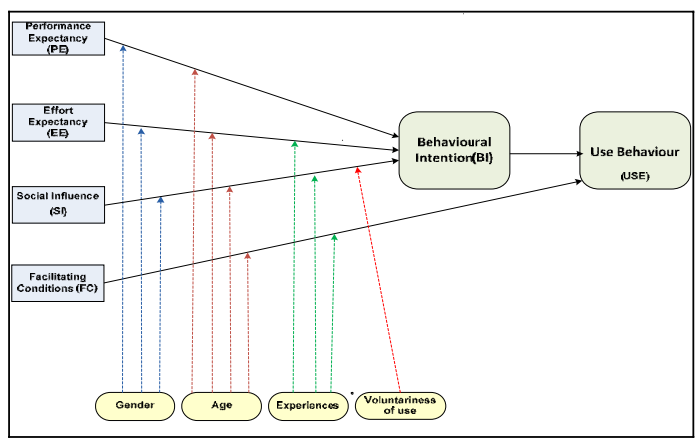

Figure 1. UTAUT model (Venkatesh et. al., 2003).

## 3. METHODLOGY

### 3.1 Research Model

The objective of this study is to identify factors that influence the intention of Saudi's citizens to use e-government services. The research model of the study is presented below in Figure 2. The constructs included in the research model were drawn from the variables used in the Venkatesh et al. (2003) study as shown earlier. In this study behavioural intention has been used to indicate the actual influence on usage of e-government services. It is mentioned in many research studies that the behavioural intentions will have a positive and direct influence on usage behavior (Venkatesh *et al.,* 2003). Also, Irani *et al.,* (2008), state that the majority of technology adoption research has utilized behaviour intention to predict technology adoption.

Furthermore, Ajzen (1991) said that behavioural intention has a direct influence on adoption of technology. In addition, the relation between behavioural intention to use a technology and actual usage is well-established (Ajzen 1991; Taylor & Todd 1995; Venkatesh & Morris 2000) and both variables could be used to measure technology acceptance. For brevity, in current study the behavioural intention to use e-government service will be used to measure the actual usage of e-government services in KSA while it is highly correlated with Use Behavior.

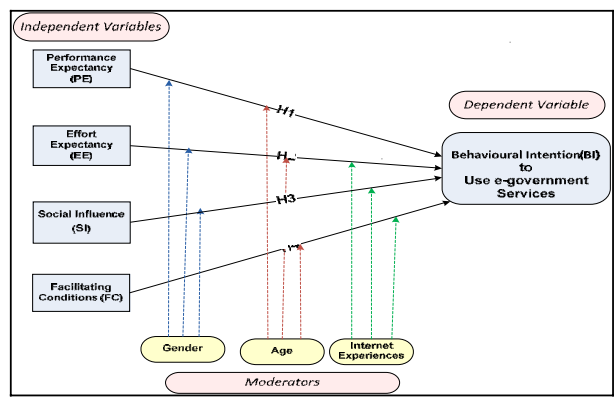

Figure 2. Research Model

## 3.2 Research Hypotheses

The researchers hypothesized relationships between variables as follow:

Table 1. Research Hypothesis

| NO. | Hypothesis |
|---|---|
| H1 | Performance expectancy will have a positive influence on behavioral intentions to use e-government services. |
| *H1a* | Gender will positively moderate the influence of performance expectancy on behavioral intentions to use e-government services for men. |
| *H1b* | Age will positively moderate the influence of performance expectancy on behavioural intentions to use e-government services. |
| H2 | Effort expectancy will have a positive influence on behavioral intentions to use e-government services. |
| *H2a* | Gender will positively moderate the influence of effort expectancy on behavioral intentions to use e-government services. |
| *H2b* | Age will positively moderate the influence of effort expectancy on behavioral intentions to use e-government services. |
| *H2c* | Internet experiences will positively moderate the influence of effort expectancy on behavioral intentions to use e-government services. |
| H3 | Social influence will have a positive influence on behavioral intentions to use e-government services. |
| *H3a* | Gender will positively moderate the influence of social influence on behavioral intentions to use e-government services. |
| *H3b* | Internet experiences will positively moderate the influence of social influence on behavioral intentions to use e-government services. |
| H4 | Facilitating conditions will have a positive influence on behavioral intentions to use e-government services. |
| *H4a* | Age will positively moderate the influence of facilitating conditions on behavioral intentions to use e-government services. |
| *H4b* | Internet experiences will positively moderate the influence of facilitating conditions on behavioral intentions to use e-government services. |

## 3.3 Data Collection

In this study quantitative research methods that include questionnaires survey were used to conduct an interpretive study. The questionnaire was divided into different sections for easy reading and completion and use a Likert scale with five levels of possible answers respect to UTAUT model (from Strongly Disagree to Strongly Agree). Questionnaires were distributed to a variety of Saudi citizens in public places such as: shopping centres, parks, hospitals, ministries and internet café. Therefore, it targeted different age groups and most likely targeted Internet users. A total of 500 questionnaires were distributed, of which 400 were completed usable for study from a variety types of Saudi citizens.

## 4. DATA ANALYSIS AND RESULT

Statistical Packages for Social Science (SPSS) was used to analyse the data collected through the surveys. The study applied the Structural equation modelling (SEM) technique to evaluate the relationships in the UTAUT model and to test the hypotheses among the variables in the model. Structural Equation Modelling (SEM) is a statistical methodology that takes a confirmatory (hypothesis testing) approach to the structural analysis of data representing some phenomena (Kline, 2005). The following section will illustrate the study analysis in more detail.

## 4.1 Descriptive Statistical Perspective

The following Table 2 provides a general overview of the Saudi citizens participated in this study in terms of the demographic information, such as gender, age and education level.

Table 2. Demographic information of respondents

| | Variable | Frequency | Percent |
|---|---|---|---|
| Gender | Male | 290 | 72.5 |
| | Female | 110 | 27.5 |
| Age | Less than 20 | 55 | 13.7 |
| | 21-30 | 195 | 48.7 |
| | 31-40 | 132 | 33.2 |
| | 41-50 | 15 | 3.7 |
| | More than 50 | 3 | 0.7 |
| Education | H.Shool | 22 | 5.5 |
| | Diploma | 168 | 42.0 |
| | Bachelor | 195 | 48.7 |
| | Higher education | 15 | 3.8 |

## 4.2 Reliability Verification

The reliability of a measure refers to the degree to which the instrument is free of random error. It is concerned with consistency and stability of the measurement. Internal consistency tends to be a frequently used type of reliability in the IS domain (Sekaran, 2003). In this study Cronbach's coefficient alphas, which are calculated based on the average inter-item correlations, were used to measure internal consistency. Reliability coefficient was run on SPSS for each set of constructs and the results are presented in Table 3. Overall, the result shows that all alpha values of the study instrument are reliable and exhibits appropriate construct reliability.

Table 3. Cronbach alpha reliability results

| Constructs | No. Of Items | Cronbach Alpha ($\alpha$) |
|---|---|---|
| Performance Expectancy (PE) | 5 | 0.83 |
| Effort Expectancy (EE) | 4 | 0.84 |
| Social Influence (SI) | 5 | 0.77 |
| Facilitating Condition (FC) | 3 | 0.83 |

## 4.3 Validity Test

Construct validity is defined as the degree to which an operational measure correlates with the theoretical concept investigated. In this study, confirmatory factor analysis was conducted to assess the overall measurement models and examine the convergent and discriminant validity.

### 4.3.1 Convergent Validity

Convergent validity is a function of the association between two different measurement scales which are supposed to measure the same concept, and is achieved when multiple indicators operate in a consistent manner (Gefen and Straub, 2005). In the confirmatory factor analysis, converge validity rely on the average variance extracted (AVE) as a base. AVE was mainly used to calculate the explanatory power of all variable of the dimension to the average variations. Constructs have convergent validity when the composite reliability exceeds the criterion of 0.70 and the average variance extracted is above 0.50 (Hair, et al., 2006). Table 4 shows that all composite reliabilities exceeded the criterion of 0.70.

Table 4. Convergent Validity

| Construct | Composite Reliability | Average Variance Extracted |
|---|---|---|
| Performance expectancy | .83 | .85 |
| Effort expectancy | .84 | .85 |
| Social influence | .77 | .79 |
| Facilitating conditions | .83 | .84 |

### 4.3.2 Discriminant Validity

Discriminant validity is the extent to which scales reflect their suggested construct differently from the relation with all other scales in the research model (Straub et al. 2004). Discriminant validity is assessed by comparing the square roots of average variance extracted (AVE) to the inter-factor correlations between constructs. Hair et al. (2006) asserted that if the AVE is higher than the squared inter-scale correlations of the construct then discriminant validity is supported. As shown in Table 5, all square roots of AVEs (diagonal cells) are higher than the correlations between constructs and that definitely confirming adequate discriminant validity.

Table 5. Discriminant validity result

| Construct | PE  | EE  | SI  | FC  |
|-----------|-----|-----|-----|-----|
| PE        | .92 |     |     |     |
| EE        | .21 | .92 |     |     |
| SI        | .55 | .16 | .89 |     |
| FC        | .29 | .50 | .38 | .88 |

## 4.4 Hypothesis Testing Results

Testing the hypotheses aims to determine which predictors (independent variables) provide a meaningful contribution to the explanation of the dependent variables (Hair et al, 2006). In this study, hypotheses testing was conducted using AMOS 19.0. Table 6 represents the results of testing the current research hypotheses. The 'Conclusion' column indicates whether that hypothesis was: supported or not supported depending on the result coefficients beta.

Table 6. Hypothesis testing results

| Hypotheses/Path | Finding | Conclusion |
|-----------------|---------|------------|
| H1(PE→BI)       | Beta= 0.34 | Supported |
| H1a             | not significant | Not supported |
| H1b             | not significant | Not supported |
| H2 (EE→BI)      | 0.54 | Supported |
| H2a             | not significant | Not supported |
| H2b             | not significant | Not supported |
| H2c             | Beta=0.43 | supported |
| H3(SI→BI)       | Beta=0.042 | Not supported |
| H3a             | not significant | Not supported |
| H3b             | Beta=0.30 | supported |
| H4(FC→BI)       | 0.38 | Supported |
| H4a             | not significant | Not supported |
| H4b             | Beta= 0.41 | supported |

**$p < 0.001$.

## 5. DISCUSSION

This section discussed the results of the survey based on the finding of hypotheses result. As shown in Table 6 the impact of the factors in the study model and its influences on adoption process can be classified to significant and non-significant factors as follow:

## 5.1 Significant Factors and Moderators

*Performance Expectancy (PE)* had a positive effect on Behaviour Intention, but there is no effect of gender or age as moderators to this relationship. This result emphasise that performance expectancy remains significant and a strong factor of behavioural intention (Venkatesh et al., 2003).

*Effort Expectancy (EE)* had a positive effect on Behavioural Intention to use e-government services, and this relationship would be moderated by Internet experience only. Age and gender were not being considered as important moderators in this connection. This result evince that effort expectancy is a significant predictor of behavioural intention (Venkatesh et al., 2003)

*Facilitating Conditions (FC)* had a positive effect on Behavioural Intention to use e-government services, and this relationship would be moderated by Internet experience only. Age and gender were not being considered as important moderator in this relationship.

○ *Internet experience* was found to be significant moderator in terms of influencing the behavioural intention to use e-government services in KSA.

## 5.2 Non Significant Factors and Moderators

● Social Influence (SI) did not have a significant effect on Behavioural Intention to use e-government services and its hypothesis has not been supported but it would be moderated by Internet experience only. According to Venkatesh *et al.*, (2008), to consider facilitating conditions as a predictor for behavior intentions, individuals' perceptions about these factor should perfectly and logically reflect their actual control over a behavior.

Age and gender were found to be insignificant in terms of moderating the behavioural intention to use e-government services in KSA.

## 6. CONCLUSIONS

This study applies presently amended UTAUT model on user acceptance and use of e-government services in KSA. Based on the data collected and the results of the analysis, it can be concluded that Performance Expectancy, Effort Expectancy, and Facilitating Condition have positive influences on user intention to use e-government services. However, in this study Social Influence was found to be insignificant in terms of predicting the behavioural intention to use e-government services and its hypothesis was not supported. In future work, we would add Trust and website Quality as independent variable into our research model and consider the effects of other crucial constructs of the UTAUT model within the context of Saudi environment. To be more precise and convincing, our work will continue and new findings will be anticipated.